\documentclass[runningheads]{llncs}
\usepackage[T1]{fontenc}
\usepackage{graphicx}
\usepackage[colorlinks=true, allcolors=blue]{hyperref}
\usepackage{color}
\usepackage[numbers,sort&compress]{natbib}
\usepackage{todonotes}

\usepackage{longtable}
\usepackage{amsmath}
\usepackage{amssymb}
\usepackage{algorithm}
\usepackage{algpseudocode}
\usepackage{listings}
\usepackage[frozencache]{minted}
\usepackage{booktabs}
\usepackage{subcaption}

\usepackage{tikz}
\usepackage{bbding} 
\usetikzlibrary{shapes,arrows}

\newcommand*{\extended}{}%
\newcommand{\ignore}[1]{}
\newcommand{\angl}[1]{\langle #1 \rangle}

\newcommand{\TD}{\textbf{TD}}
\newcommand{\TTD}{\textbf{TLTD}}

\newcommand{\get}{{\textsf{get}}}
\newcommand{\set}{{\textsf{set}}}
\newcommand{\main}{{\textsf{main}}}
\newcommand{\demand}{{\textsf{demand}}}
\newcommand{\Put}{{\textsf{foo}}}
\newcommand{\eq}{{\textsf{eq}}}

\newcommand{\I}{{\mathbb{I}}}
\newcommand{\wizwoz}{{\textsc{Goblint}}}
\newcommand{\imme}{\emph{immediate}}
\newcommand{\indep}{\emph{independent}}
\begin{document}
\title{Same Engine, Multiple Gears:
	Parallelizing Fixpoint Iteration at Different Granularities}
\ifdefined\extended
\subtitle{(Extended Version)}
\fi
\titlerunning{Parallelizing Fixpoint Iteration}
%Parallelizing a generic fixpoint engine}
% \ignore{
\author{Ali Rasim Kocal\inst{1}\orcidID{0009-0004-9019-0852} \Envelope\and
Michael Schwarz\inst{2}\orcidID{0000-0002-9828-0308} \and
Simmo Saan\inst{3}\orcidID{0000-0003-4553-1350} \and
Helmut Seidl\inst{1}\orcidID{0000-0002-2135-1593}}
\authorrunning{A. R. Kocal et al.}
\institute{Technical University of Munich, Garching, Germany\and
National University of Singapore, Singapore\and
University of Tartu, Estonia\\
\email{\{ali.rasim.kocal,helmut.seidl\}@tum.de\\
m.schwarz@nus.edu.sg, simmo.saan@ut.ee}
}
% }

%
\maketitle              % typeset the header of the contribution
\newcommand{\numberConratBenches}{13}
\newcommand{\numberPthreadBenches}{4}
\newcommand{\numberSvBenches}{30}
\newcommand{\numberTotalBenches}{47}
\newcommand{\visualSamplingFactor}{fourth}
\newcommand{\refAvg}{1.18}
\newcommand{\baseOneMin}{0.6}
\newcommand{\baseOneAvg}{2.25}
\newcommand{\baseOneMax}{10.42}
\newcommand{\baseFuncOneMin}{0.41}
\newcommand{\baseFuncOneAvg}{1.85}
\newcommand{\baseFuncOneMax}{9.49}
\newcommand{\baseTwoSelfRelativeMin}{0.75}
\newcommand{\baseTwoSelfRelativeAvg}{1.7}
\newcommand{\baseTwoSelfRelativeMax}{4.98}
\newcommand{\baseFourSelfRelativeMin}{0.54}
\newcommand{\baseFourSelfRelativeAvg}{2.01}
\newcommand{\baseFourSelfRelativeMax}{6.58}
\newcommand{\baseEightSelfRelativeMin}{0.36}
\newcommand{\baseEightSelfRelativeAvg}{2.68}
\newcommand{\baseEightSelfRelativeMax}{9.26}
\newcommand{\baseFuncTwoSelfRelativeMin}{0.52}
\newcommand{\baseFuncTwoSelfRelativeAvg}{1.8}
\newcommand{\baseFuncTwoSelfRelativeMax}{5.03}
\newcommand{\baseFuncFourSelfRelativeMin}{0.71}
\newcommand{\baseFuncFourSelfRelativeAvg}{2.44}
\newcommand{\baseFuncFourSelfRelativeMax}{11.29}
\newcommand{\baseFuncEightSelfRelativeMin}{0.21}
\newcommand{\baseFuncEightSelfRelativeAvg}{3.18}
\newcommand{\baseFuncEightSelfRelativeMax}{9.98}
\newcommand{\distTwoSelfRelativeMin}{0.79}
\newcommand{\distTwoSelfRelativeAvg}{1.83}
\newcommand{\distTwoSelfRelativeMax}{4.25}
\newcommand{\distFourSelfRelativeMin}{0.73}
\newcommand{\distFourSelfRelativeAvg}{2.5}
\newcommand{\distFourSelfRelativeMax}{7.71}
\newcommand{\distEightSelfRelativeMin}{0.42}
\newcommand{\distEightSelfRelativeAvg}{2.87}
\newcommand{\distEightSelfRelativeMax}{9.5}
\newcommand{\baseSolverWorsePrecisionRate}{0.03}
\newcommand{\baseSolverBetterPrecisionRate}{0.04}
\newcommand{\baseSolverEqualPrecisionRate}{0.92}
\newcommand{\baseSolverPerfectionRate}{0.601063829787234}
\newcommand{\baseSolverPrecisionEqualOrMoreRatio}{0.97}
\newcommand{\baseSolverPracticallyIdenticalRate}{0.9148936170212766}
\newcommand{\baseFuncSolverWorsePrecisionRate}{0.03}
\newcommand{\baseFuncSolverBetterPrecisionRate}{0.04}
\newcommand{\baseFuncSolverEqualPrecisionRate}{0.92}
\newcommand{\baseFuncSolverPerfectionRate}{0.5585106382978723}
\newcommand{\baseFuncSolverPrecisionEqualOrMoreRatio}{0.97}
\newcommand{\baseFuncSolverPracticallyIdenticalRate}{0.9148936170212766}
\newcommand{\distSolverWorsePrecisionRate}{0.05}
\newcommand{\distSolverBetterPrecisionRate}{0.06}
\newcommand{\distSolverEqualPrecisionRate}{0.89}
\newcommand{\distSolverPerfectionRate}{0.5691489361702128}
\newcommand{\distSolverPrecisionEqualOrMoreRatio}{0.95}
\newcommand{\distSolverPracticallyIdenticalRate}{0.723404255319149}

\begin{abstract}
Fixpoint iteration constitutes the algorithmic core of static analyzers.
Parallelizing the fixpoint engine can significantly reduce analysis times.
Previous approaches typically fix the granularity of tasks upfront, e.g., at the
level of program threads or procedures --- yielding an engine permanently stuck in one gear.
Instead, we propose to parallelize a generic fixpoint engine in a way that is parametric in the
task granularity --- meaning that our engine can be run in different gears.
We build on the top-down solver \TD, extended with support for mixed-flow sensitivity,
and realize two competing philosophies for parallelization, both building on a task pool that schedules
tasks to a fixed number of workers. The nature of tasks differs between the philosophies.
In the \imme\ approach, all tasks access a single thread-safe hash table maintaining
solver state, while in the \indep\ approach, each task has its own state
and exchanges data with other tasks via a \emph{publish/subscribe} data structure.
We have equipped the fixpoint engine of
the static analysis framework
\wizwoz\ with implementations following both philosophies
and report on our results for large real-world programs.

\keywords{abstract interpretation  \and static analysis \and fixpoint iteration \and parallelization.}
\end{abstract}

\section{Introduction}\label{s:introduction}

Static analysis based on abstract interpretation tries to infer program invariants without running the program.
For that, candidate invariants are represented by \emph{abstract values} from some bounded lattice (the \emph{domain})
where the lattice operations \emph{meet} and \emph{join} correspond to logical conjunction and disjunction
respectively.
The analysis of the program then gives rise to a system of equations over the domain whose solutions
correspond to inductive invariants of the program.
% Why generic
The design of such an analysis consists of defining and implementing the abstract domains and the
corresponding abstract transfer functions corresponding to edges in control-flow graphs,
as well as specifying the relationships among abstract values of, e.g.,
program points (in context) or global variables.

Ideally, the designer of such an analysis should concentrate on these matters
without simultaneously being concerned with the algorithmic details of how a solution of the
system of equations is computed.
A separation of the design of the analysis and the implementation of the fixpoint engine is important --
particularly when the fixpoint engine is intricate.
As a result, fixpoint engines such as~\cite{SeidlV21,spi} are ideally implemented in an
\emph{application-independent} manner to enable re-use.
\TD~\cite{SeidlV21} is one such fixpoint engine. It is broadly applicable as it makes no assumptions on
the finiteness of the set of unknowns, requires no a priori knowledge about
dependencies between variables of the equation systems (which we call \emph{unknowns} to distinguish them from
program variables), and dynamically detects points where to apply widening/narrowing.
Moreover, it supports \emph{mixed-flow sensitive} analyses
where some unknowns are analyzed flow-insensitively while the values of others are determined flow-sensitively~\cite{Stemmler25}.

While analyzers such as \textsc{Astrée}~\cite{astree} and \textsc{Pikos}~\cite{pikos} have been parallelized,
many others, such as \wizwoz~\cite{Goblint2016}, are sequential programs and thus do not
exploit multiple cores for speeding up the fixpoint computation.
Here, we aim to provide \emph{generic} parallelization strategies that do not pose
restrictions on the programs that can be analyzed, and can easily be
adapted to a variety of analyses, in particular, mixed-flow sensitive analyses,
and support multiple \emph{granularities} of tasks.
Rather than supporting only a predefined set of granularities,
tasks are identified by \emph{root} unknowns.
For each root $x$, a separate
instance of the solver is created for locally solving for $x$ by exploring the equation system
in a demand-driven way.
To realize a parallelization of the analysis, e.g., at the level of program threads,
the endpoints of created program threads are selected as roots.
For parallelizing the analysis of sequential code, potential choices are the endpoints of procedures
or branches of specific  \emph{if-else} statements.
To minimize the synchronization overhead at accesses to shared data,
distinct roots ideally should result in the exploration of disjoint sets of unknowns.
In practice, though, this can hardly be guaranteed.
We propose two approaches for dealing with the interaction between distinct solver tasks:
\begin{itemize}
\item	The \imme\ approach maintains all solver data, grouped by
	the unknowns to which they refer, within one shared thread-safe data structure, which
	guarantees \emph{transactional} accesses by distinct tasks to individual unknowns.
\item	The \indep\ approach, on the other hand, maintains multiple local copies of solver data
	and introduces a \emph{publish-subscribe} mechanism to share information between different
	tasks.
\end{itemize}
The rest of this paper is structured as follows:
Section \ref{s:background} describes by example how \emph{side-effecting equation systems}
can be used to formalize the analysis of multi-threaded C programs.
Section \ref{s:solver} introduces the improved version \TTD\ (toplevel \TD) of \TD\ that lays the groundwork for our parallelization.
Section \ref{s:method} then presents our two approaches for parallelizing \TTD, which are experimentally evaluated
in Section \ref{s:results}.
After discussing related work on parallel solvers in Section \ref{s:work},
we summarize our results in Section \ref{s:final}.

\lstset{numbers=left}

\section{Background}\label{s:background}

The basic setup of thread-modular analysis, as considered, e.g., in \cite{schwarz2023clustered,schwarzthesis},
aims at inferring a single invariant for each program point.
To that end, for each program point $v$, an \emph{unknown} is introduced taking values from
some bounded lattice $\mathbb D$ of candidate invariants.
A right-hand side function $f_v$ is provided for every program point $v$
to indicate how the invariant for $v$ can be obtained from the invariants for other
program points, while taking the control-flow of the program into account.
Such unknowns are called \emph{locals}.
In a refined setting, multiple invariants per program point may be computed, differentiated, e.g.,
by calling-contexts or thread \emph{id}s (or both). In that case, multiple locals may be introduced for
each program point, e.g., the unknown $\angl{v,c}$ for program point $v$ in context $c$.
To abstract thread interaction via shared data structures, a dedicated further set of \emph{globals} is introduced.
These are unknowns whose values represent invariants of shared data that are meant to hold throughout the whole
program execution.
Their values are therefore accumulated \emph{flow-insensitively}.

\begin{figure}[hbt]
\begin{center}\begin{minipage}{8cm}
\begin{minted}[linenos,escapeinside=||]{c}
int g;
void* foo(void* a) {
    g = (int) a; |\label{f:running:gh}|
    return NULL;
}
int main() {
    pthread_t t1; int a;
    g = 0;
    pthread_create(&t1, NULL, foo, (void *) 42); |\label{f:running:create}|
    a = g;
    a = a+1; |\label{f:running:hpp}|
    return a;
}
\end{minted}
\end{minipage}
  \caption{\label{f:running}A multi-threaded C program}
\end{center}
\end{figure}
Consider the multi-threaded C program in Figure \ref{f:running}, which serves as our running example.
Assume that we are going to perform \emph{interval} analysis~\cite{popl77} in a thread-modular way.
Interval analysis for the program in Figure \ref{f:running} attempts to infer tight lower and upper bounds for the values of $a$ and $g$.
Since the variable $g$ is shared, we introduce a global unknown for $g$, for convenience, with the same name.
As a bounded lattice of candidate invariants for $g$,
we use the set $\I$ of intervals (extended with $\emptyset$) equipped with the natural ordering.

The set of locals in our examples consists of the set of program points.
For technical reasons, we consider \emph{start points} of called procedures as globals,
since these do not receive their values via right-hand sides, but via contributions from
call-sites dispersed all over the program.
In general, interval analysis will maintain a separate interval for each local program variable.
In our minimalistic example, however, there is just a single local variable $a$ in both functions \textsf{main}
and \textsf{foo}.
The candidate invariants for locals may thus be taken from the same domain $\I$ as for globals in this case.
In the following, we assume that locals are named according to lines in the program text of Figure \ref{f:running}.

\section{The Solver \TTD}\label{s:solver}

Let $\mathcal X$ denote the set of all unknowns.
A run of the analysis consists of repeatedly evaluating right-hand sides until a
valuation of unknowns $\sigma:\mathcal X\to\mathbb D$ is found so that each right-hand side
is accounted for. Specifically, this means that re-evaluating each right-hand side
only produces results already subsumed by $\sigma$.

In this section, we describe a version \TTD\ of the top-down solver \TD~\cite{SeidlV21}, which
has a modified organization of its data structures and an additional top-level workset.
This modification, while also beneficial on its own, is the groundwork for our parallelization.
The implementation is presented in
Figures \ref{f:td:mainloop}, \ref{f:td:core}, and \ref{f:td:destab}, where the parts differing from \TD\ are highlighted, apart from those related to the modification of data structures.

\TD\ solves for unknowns recursively as soon as they are encountered during the analysis.
In contrast, \TTD\ introduces a main loop (Figure \ref{f:td:mainloop}) that calls the function \texttt{iterate}
for each unknown $x$ in the workset \verb!work!
to determine the value for $x$.
Initially, \verb!work! is assumed to contain the \emph{unknowns of interest},
typically, the endpoint of \verb|main| ($\angl{13}$ in our example),
which triggers the iteration on all possibly \emph{influencing} unknowns.
The set \verb!work! may \emph{dynamically} receive further root unknowns
during the analysis, enabling us to solve for unknowns in parallel.

\begin{figure}
\centering
\begin{minipage}{6cm}
  \begin{minted}[linenos,firstnumber=46,highlightlines={46-48}]{ocaml}
while not (Set.is_empty !work) do (
  iterate (Set.choose !work)
) done
\end{minted}
\end{minipage}
\caption{\label{f:td:mainloop}The top-level loop of \TTD}
\end{figure}
The core of the \TTD\ algorithm is the \verb!iterate! function (Figure \ref{f:td:core}).
When called for an unknown $x$,
\verb!iterate! repeatedly evaluates the right-hand side for $x$
(line \ref{f:td:core:eq}), until a fixpoint is reached.
The right-hand side provided by the call \verb!eq!~$x$
is a higher-order function that
formalizes how the abstract value of $x$ can be determined. It is
parametrized by the three functional arguments $\get$, $\set$ and
$\demand$, which represent the \emph{interface} to the fixpoint engine.
The first functional argument $\get$ is used to access the values of other unknowns.

\begin{figure}
\centering
\begin{minipage}{10.5cm}
  \begin{minted}[linenos,escapeinside=||,highlightlines={17-20,32}]{ocaml}
let rec iterate x =
  let rec query y =
    let dy = find data y in
    if (not (dy.called) && not (is_global y)) then (
	iterate y
      );
    dy.influences <- (Set.add x dy.influences);
    dy.value in
  let side g contribution = 
    let dg = find data g in
    dg.stable <- true;
    let new_value = Domain.join dg.value contribution in
    if not (Domain.leq new_value dg.value) then (
      dg.value <- new_value;
      destabilize g;
    ) in
  let promote y = 
    let dy = find data y in
    dy.toplevel <- true;
    work := Set.add y !work in

  let dx = find data x in
  dx.called <- true;
  while not dx.stable do (
    dx.stable <- true;
    let new_value = eq x query side promote in |\label{f:td:core:eq}|
    if not (Domain.leq dx.value new_value) then (
      dx.value <- new_value;
      destabilize x;
    );
  ) done;
  if dx.toplevel then (work := Set.remove x !work);
  dx.called <- false
  \end{minted}
  \end{minipage}
  \caption{\label{f:td:core}The core fixpoint iterator of \TTD.}
\end{figure}

\begin{example}\label{e:assign}

Consider the assignment \verb!a = a + 1;! in line~\ref{f:running:hpp} of Figure \ref{f:running}.
  Then the right-hand side of the unknown $\angl{12}$ following that line is given by
\begin{equation}
\eq\;\angl{12} = \textbf{fun}\  \get\ \set\ \demand\ \rightarrow (\get\,\angl{11})\,+^\sharp\,[1,1]
  \label{eq:get1}
\end{equation}
where the $+^\sharp$ is the abstract version of the programming language operator ``+'' for the domain
$\I$.
$\hfill\qed$
\end{example}
The values for program globals are analyzed \emph{flow-insensitively} and represented by \emph{global}
unknowns (\emph{globals}) in $\mathcal{X}$. Globals do not have right-hand sides on their own, but receive contributions
from the right-hand sides of locals.
The second argument $\set$, therefore, provides a means to produce such contributions (called \emph{side-effects}).
\begin{example}
Consider the assignment \verb!g = (int) a;! in line~\ref{f:running:gh}.
Then
\begin{equation}
\eq\;\angl{4} = \textbf{fun}\ \get\ \set\ \demand\ \rightarrow
	\begin{array}[t]{l}
	\textbf{let}\,d = \get\,\angl{3}\;\textbf{in}	\\
	\set\,g\,d; \\
	d
	\end{array}
  \label{eq:set1}
\end{equation}
Here, first the value $d$ of \verb!a! is obtained via \get, which also happens to be the
value to be propagated to the unknown $\angl{4}$.
Additionally, however, a contribution is provided to the global $g$ by means of the call
$\set\,g\,d$.
$\hfill\qed$
\end{example}
The third argument \demand\ is introduced by \TTD. It can be called on unknowns,
of which the values must be calculated, e.g., for their side-effects, but are not used directly.
The following example illustrates a potential usage.

\begin{example}\label{e:create}
Consider the right-hand side for the program point 10, following the creation of a new thread
in line~\ref{f:running:create}. The created thread will concurrently execute the function \Put\ for input 42.
The corresponding equation therefore contributes the interval [42,42] as an abstraction of 42 to the
start point of \Put. The result of the evaluation of \Put\ is not required to determine the value for the
next program point 10. Nevertheless, the body of \Put\ must be analyzed, as it might affect global unknowns
through \set. Therefore, \demand\ is called for the endpoint $\angl{5}$ of \Put.
Accordingly, we define
\begin{equation}
\eq\;\angl{10} = \textbf{fun}\ \get\ \set\ \demand \rightarrow
	\begin{array}[t]{l}
	\textbf{let}\,d = \get\,\angl{9}\;\textbf{in}	\\
	\set\,\angl{3}\,[42,42]; 	\\
	\demand\,\angl{5}; \\
	d
	\end{array}
  \label{eq:demand}
\end{equation}
% ugly hack, \qedhere would be better
\hfill$\smash{\raisebox{3ex}{\qed}}$
\vspace*{-3ex}
\end{example}
To summarize, the three argument functions \get, \set\ and \demand\ represent the generic
\emph{interface} which allows \TTD\ to interact with the evaluation of right-hand sides
in the following way:
\begin{itemize}
\item	The functional parameter \get\ receives the local solver function \verb!query!.
	That function, when called for some unknown $y$, ultimately returns the value of $y$ as
	stored in \verb!data!.
	Before that, however, an up-to-date value of $y$ is calculated by calling \verb!iterate! whenever an iteration on $y$ has not
	already been started. To check the latter, \verb|data| for $y$ maintains a flag \verb|called|
	which is set by the function \verb|iterate| before it enters the \emph{while} loop to perform
	fixpoint iteration on $y$.
	Once iteration for $y$ has terminated, the dependence of $x$ on $y$ is recorded by inserting $x$
	into the set attribute \verb|influences| of $y$ in \verb|data|, and the latest value of $y$ is returned.
\item	The functional parameter \set\ receives the local function \verb!side!,
        which has arguments a global $g$ and a contribution to $g$.
	First, the attribute \verb|stable| of $g$ in \verb|data| is set to \verb|true| to
	include $g$ in the result set of unknowns.
	If the new contribution is subsumed by the old value of $g$, the solver proceeds.
	Otherwise, the value of $g$ in \verb|data| is updated to the join of old and new values,
	and \verb|destabilize| is called to inform all other unknowns that directly or indirectly
	depend on $g$, as their current values may be outdated.
\item	Finally, the solver function \verb!promote! is provided for the functional parameter \demand.
        When called for an unknown $y$, \verb!promote! adds $y$ to the top-level workset
	and records in \verb|data| that $y$ has now been raised to the top-level.
	Note that no influence is recorded for $y$.
\end{itemize}
When called on a hashmap \verb|data| and an unknown $x$, the helper function \verb|find|
-- which is left out for brevity --
looks up a record for $x$ in \verb|data|, creates one if there is none present and returns a reference
to the record consisting of the fields accessed in Figure \ref{f:td:core}. This is different
from \TD\ as presented in \cite{SeidlV21}, which organizes the data in individual hashmaps for
each field.

In detail, the iteration on $x$ performed by the function \verb|iterate| proceeds as follows:
First, the flag \verb|called| for $x$ in \verb|data| is set to \verb|true|.
The iteration on $x$ is repeated until the value of $x$ as stored in \verb|data| for $x$
does not change anymore, i.e., \emph{stabilizes}. To indicate that fact, the record \verb|data|
for $x$ has an entry \verb|stable|, which is optimistically set to \verb|true|
before the \textit{while} loop.
Whenever the evaluation of the right-hand side for $x$ provides a value that is not subsumed
by the old value, the entry \verb|value| of the record \verb|data| $x$ is updated to
the join of the old value for $x$ with the new value.
Furthermore, for all other unknowns that are directly or indirectly influenced by $x$, the entries
\verb|stable| are set to \verb|false| by calling the auxiliary function \verb|destabilize|.
The function \emph{destabilize}, as presented in Figure~\ref{f:td:destab}, additionally
checks whether a destabilized unknown $y$ is top-level. If this is the case, it is added
back to the toplevel workset \verb|work| for re-iteration, ensuring all influences are accounted for
even when $y$ is not queried again.
Finally, assume that the unknown $x$ has \emph{not} been destabilized
-- either during the last evaluation of its right-hand side or
by the last call to \verb|destabilize| after a modification of the entry \verb|value| of $x$.
Then the iteration has terminated. In case
$x$ has been marked as \verb|toplevel|, $x$ is also removed from \verb|work|.
In the end, the entry \verb|called| in \verb|data| $x$ is reset to \verb|false|.

\begin{figure}
\centering
\begin{minipage}{7cm}
  \begin{minted}[linenos,firstnumber=34,highlightlines={41-43}]{ocaml}
and destabilize x =
  let dx = find data x in
  let influences = dx.influences in
  dx.influences <- Set.empty;
  Set.iter (fun y ->
    let dy = find data y in
    dy.stable <- false;
    if dy.toplevel then (
      work := Set.add y !work
    );
    destabilize y
  ) influences
  \end{minted}
  \end{minipage}
  \caption{\label{f:td:destab}The auxiliary function \texttt{destabilize}}
\end{figure}

\begin{example}\label{e:example-run}
In our example, the global $g$ receives a contribution $[0,0]$ via the right-hand side of
program point $\angl{9}$ in $\main$.
That value will be read when evaluating the right-hand side for $\angl{11}$, and will be incremented
in the right-hand side for $\angl{12}$ and thus provides the first value for the endpoint $\angl{13}$
of $\main$.
In the right-hand side for program point $\angl{10}$, however, a contribution of $[42,42]$ occurs to
the start point $\angl{3}$ of function $\Put$, while in the same right-hand side the endpoint
$\angl{5}$ of $\Put$ is turned into a top-level unknown and inserted into \verb|work|.
The subsequent iteration on $\angl{5}$ results in a further contribution of $[42,42]$ to
$g$. By joining this new contribution with the old value of $g$, the solver obtains the interval
$[0,42]$ for $g$. The subsequent call to \verb|destabilize| will put the endpoint $\angl{13}$
of $\main$ back into the workset \verb|work|.
In a second go on the body of $\main$, the solver will recompute the values for
$\angl{11}$ and $\angl{12}$ to $[0,42]$ and $[1,43]$, respectively, which
then is also the final value for the endpoint of $\main$.
$\hfill\qed$
\end{example}
The changes of \TTD\ compared to \TD\ prepare the ground for parallelization:
\begin{itemize}
\item 	Fusing several data structures (as maintained by \TD) into one
	reduces the number of possibly expensive accesses to large data structures,
	which later may be shared between multiple tasks.
\item	Introducing a top-level workset provides a mechanism to replace the recursive descent
	into unknowns with a \emph{delayed} or, as described in Section \ref{s:method},
	\emph{parallel} execution.
\end{itemize}
The first modification does not change the semantics of the solver compared to the original \TD\ 
(a formal correctness proof of the original \TD\ without side effects is given at \cite{partialcav,partial}).
The second modification assures that each delayed unknown $x$ once added to the top-level
worklist, will be re-entered into that worklist whenever it is destabilized.
In this way, $x$ is guaranteed to be iterated upon until stabilization.

The version of \TTD\ as implemented in \wizwoz\ additionally supports
\emph{widening} and \emph{narrowing}~\cite{popl77,cousot1992}.
This addition is orthogonal to its parallelization. Therefore, it has been omitted here.

\section{Parallelizing \TTD}\label{s:parallel}\label{s:method}

Task-pool-based parallelization is a common approach, in which a number of \emph{tasks} are scheduled on a generally lower
number of processors. It has, e.g., been used for the parallelization
of \emph{modular} analyses~\cite{modularworklist}.
This approach is suitable when the analysis consists of sufficiently large modules, such as threads or functions.
However, for solving systems of equations as described in Section \ref{s:introduction},
the unknowns, and therefore tasks, would correspond not to large components but to individual program points.
With such tiny components in large numbers,
creating one solver thread for every unknown is impractical.
Our idea is to use only the top-level workset of \TTD\ as our task pool for parallelization.
Tasks are identified dynamically by root unknowns triggering the \emph{recursive exploration} of larger parts
of the equation system, and are mapped to a fixed set of workers by the task-pool.
This enables parallelization without a priori knowledge of the equation system.
However, the partitioning into solver tasks takes place in a non-deterministic fashion, and the right-hand sides may be evaluated in different orders in multiple solver runs.
In the presence of \emph{widening and narrowing}, this can lead to solutions with varying precision.
In our running example, the top-level workset is initialized with the endpoint of $\main$ and the endpoint of $\Put$ later makes it into the top-level workset.
Calling \verb|iterate| for these in different tasks will trigger one task to explore the unknowns from
the body of $\main$ and further unknowns from the body of $\Put$.
We propose two different approaches to how the interaction between multiple tasks is organized.
\begin{itemize}
\item   The \imme\ approach maintains all solver data within one shared thread-safe data structure, which
	guarantees \emph{transactional} accesses by distinct workers to individual unknowns.
\item   The \indep\ approach, on the other hand, maintains multiple local copies of solver data
        and introduces a \emph{publish-subscribe} mechanism to share information between tasks.
\end{itemize}
The \imme\ approach assures that at any point in time, at most one worker is iterating on any given unknown.
In this way, spurious computation is ruled out -- at the price of synchronization at every data access.
The \indep\ approach does not attempt to exclude duplication of work.
Instead, it reduces the need for synchronization to the accesses to the publish-subscribe mechanism.
In detail, these approaches work as follows.

\subsection{The \textit{Immediate} Approach}\label{ss:helmut}	% Helmut
The \imme\ approach relies on a thread-safe implementation of the data structure \verb|data|.
To ensure that a worker operating on an unknown $x$ only writes consistent results to
\verb|data|, i.e., results that are based on the latest values in \verb|data|, we use a
\emph{transactional} model (see, e.g., \cite{DBLP:journals/jpdc/Grahn10}).
Thus, for any operation on $x$, such as \verb|iterate|,
the data is read at the beginning of the operation.

If the data changes before results are saved in the shared data, we repeat the whole operation with newly read data.
In the implementation, this is achieved by replacing the records in \verb|data| with atomic records
on which we can apply \emph{compare-and-swap} operations (CAS). The CAS operation sets the new value of a
record only if the last known value, as provided by a second parameter, is unchanged.
Note that this approach enables a \emph{lock-free} implementation thanks to the fact that \TTD\ keeps a single reference for all data related to an unknown. We use a hashmap with buckets implemented as \emph{Michael-Scott} queues~\cite{DBLP:conf/podc/MichaelS96}.
In our application, the atomic wrapper for the data of an unknown, once initialized, never changes -- only the values
it refers to may be updated. Moreover, it is never removed from
\verb|data|, allowing for a simplification and efficient implementation of the hashmap.

For good performance, repeated operations due to CAS-failures must be exceedingly rare.
Note that once a task $t$ has called the function \verb|iterate| on an unknown $x$, no other task
will call \verb|iterate| on $x$ until $t$ finishes iteration and sets \verb|called| to \verb|false|.
Therefore, $t$ has sole write access to the entries \verb|called| and \verb|value| of $x$,
whereas other solver threads only read them.
The only field that is updated frequently by other solver threads,
potentially causing an undue amount of CAS failures, is \verb|influences|.
Therefore, \verb|influences| is implemented as a \emph{thread-safe set} and excluded from the transactional model.
This design decision reduces the ratio of repeated operations due to CAS failures down to less than 0.01\% on average.
To maintain correctness, nonetheless, it must be ensured that every unknown $y$ that depends on $x$ is
destabilized when $x$ receives a new value. We achieve this by always destabilizing $x$ \emph{after}
a new value is saved
and expanding \verb|influences| of $x$ with $y$ \emph{before} the value of $x$ is read during the iteration for $y$.
When $x$ receives a new value between the expansion of \verb|influences| and the reading of the new value,
this strategy can lead to premature destabilization of $y$ and an unnecessary recalculation resulting
in the same value. However, this is a rare case with minimal impact on performance.

The solver terminates once the iteration for all encountered unknowns marked as \emph{toplevel} has terminated.

\subsection{The \textit{Independent} Approach}\label{ss:michael}	% Michael

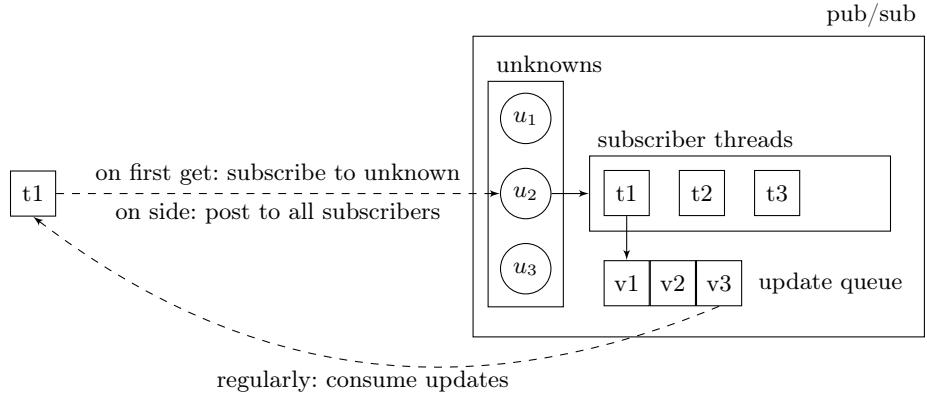
\begin{figure}[t]
\centering
\begin{tikzpicture}[auto, node distance=2cm,>=latex']
	% Large rectangular block in the middle
	\node [draw, rectangle, minimum width=6cm, minimum height=4cm] (pubsub) {};

	% 'unknowns' rectangle to the right of 'popsub'
	\node [draw, rectangle, minimum height=3cm, minimum width=1cm, anchor=north west, xshift=2mm, yshift=-6mm] at (pubsub.north west) (unknowns) {};
	% three circles labeled u_1, u_2, u_3 inside 'unknowns'
	\node [draw, circle, minimum size=6mm] (u1) at ([xshift=0mm, yshift=-5mm] unknowns.north) {$u_1$};
	\node [draw, circle, minimum size=6mm] (u2) at ([xshift=0mm, yshift=-15mm] unknowns.north) {$u_2$};
	\node [draw, circle, minimum size=6mm] (u3) at ([xshift=0mm, yshift=-25mm] unknowns.north) {$u_3$};

	\node [draw, rectangle, minimum height=6mm, minimum width=6mm, anchor=east, xshift=-59mm] at (u2.west) (t1) {t1};

	\draw [->, dashed] (t1.east) -- (u2.west) node [midway, above] {on first get: subscribe to unknown} node[midway, below] {on side: post
	to all subscribers};

	% horizontal box to the right of the second circle
	\node [draw, rectangle, minimum height=1cm, minimum width=4cm, anchor=west] at ([xshift=5mm] u2.east) (subs) {};

	% arrow from u2 to subs
	\draw [->] (u2.east) -- (subs.west);

	% three small boxes inside subs
	\node [draw, rectangle, minimum height=6mm, minimum width=6mm] (s1) at ([xshift=5mm, yshift=-5mm] subs.north west) {t1};
	\node [draw, rectangle, minimum height=6mm, minimum width=6mm] (s2) at ([xshift=15mm, yshift=-5mm] subs.north west) {t2};
	\node [draw, rectangle, minimum height=6mm, minimum width=6mm] (s3) at ([xshift=25mm, yshift=-5mm] subs.north west) {t3};

	% under subs, three horizontally connected small boxes
	\node [draw, rectangle, minimum height=6mm, minimum width=6mm] (p1) at ([xshift=5mm, yshift=-17mm] subs.north west) {v1};
	\node [draw, rectangle, minimum height=6mm, minimum width=6mm, anchor=west] (p2) at (p1.east) {v2};
	\node [draw, rectangle, minimum height=6mm, minimum width=6mm, anchor=west] (p3) at (p2.east) {v3};

	% update queue label right to v3
	\node [anchor=west] at ([xshift=1mm] p3.east) {update queue};

	\draw [->] (s1.south) -- (p1.north);

	% Curved arrow from p3 to t1 with label "consume updates" below
	\draw [->, bend left, dashed] (p3.south) to node [midway, below] {regularly: consume updates} (t1.south);
	\node [anchor=south west] at (unknowns.north west) {unknowns};
	\node [anchor=south east] at (pubsub.north east) {pub/sub};
	\node [anchor=south west] at (subs.north west) {subscriber threads};
\end{tikzpicture}
	\caption{Pub/Sub system for global unknown updates}
	\label{f:pubsub}
\end{figure}

Ideally, a good selection of root-unknowns leads to recursive descent into mostly distinct parts of the
equation system. Notably, this is the case when root unknowns correspond to the endpoints of program threads.
Our second approach, therefore, assumes that, if at all, few local unknowns will be encountered by multiple tasks,
and their recalculation is less costly than the synchronization overhead.
Thus, it lets each task operate on a \emph{distinct copy} of the hashmap \texttt{data}.
Synchronization is only necessary on the values of globals as
they do not have right-hand sides, but may receive contributions from multiple locals.
We use a publish/subscribe system \cite{pubsub} to share those contributions among tasks, as illustrated in Figure \ref{f:pubsub}.
Upon first read of a global $g$, a worker \emph{subscribes} to the contributions for $g$,
while each contribution to $g$ is \emph{published} to its subscribers' thread-safe update-queues.
This is easily achieved by extending the \texttt{query} and \texttt{side} functions.
Additionally, all contributions for globals are kept, so that new subscribers can
receive the most up-to-date values as well.
Each task regularly (e.g., after the evaluation of every right-hand side)
consumes all new updates to the globals it has subscribed to.
It does so by updating their values in its local copy of \verb!data! and destabilizing influenced unknowns
if necessary.
Further, tasks are revived if a global unknown that they have subscribed to
receives an update after the task has terminated, possibly invalidating its former results.
This guarantees that all the relevant updates to the globals are taken into account by every solver task.
The solver then terminates when all tasks have terminated, and all update-queues are fully processed.
Apart from the handling of the globals, the termination criteria ensuring correctness are the same as in \TTD.

\newcommand{\minLoc}{23}
\newcommand{\maxLoc}{12.5k}
\section{Experimental Results}\label{s:results}
%%%
We have implemented both philosophies for parallelizing \TTD\ in \textsc{OCaml 5.3} using the module \textsc{Domains}
for obtaining true concurrency~\cite{ocaml5-parallel}.
The improved \emph{sequential} solver \TTD\ from Section~\ref{s:background} (extended with widening and narrowing)
can be recovered by running the \imme\ version of \TD\ with a single worker thread.
These solvers have been integrated into the open-source static analyzer \wizwoz\ for multi-threaded C.

The downside of parallelization is that, as the \emph{running time} and \emph{precision}
of the result may depend on the \emph{order} in which right-hand sides are evaluated,
the behavior may differ from one parallel run to another, though all returned results are
guaranteed to be sound.

Our goal, therefore, is to answer the following Research Questions.
\begin{description}
    \item[\textbf{RQ1}.]
    	What is the impact of the improvements from Section~\ref{s:background}
	already on the \emph{sequential} performance of \TTD?
    \item[\textbf{RQ2}.]
    	How do the \imme\ and \indep\ approaches compare to each other
	and scale with the number of worker threads?
    \item[\textbf{RQ3}.]
     	How does the choice of granularity, i.e., the
	placement of demands, affect the efficiency of the parallelization?
    \item[\textbf{RQ4}.]
    	How does the precision vary for the two approaches and between different runs?
\end{description}
To answer these questions, we configured \wizwoz\ to run a
context-sensitive interprocedural analysis with some loop unrolling.
The analyzer computes points-to information for pointers, and interval sets for \emph{int} variables,
while tracking symbolic thread \emph{id}s \cite{schwarz2023clustered} and sets of definitely held mutexes.

We have collected benchmark programs from the \emph{Concurrency} suite of SV-Comp \cite{SVCOMP25} and
the Concrat~\cite{concrat} benchmarks.
We have selected the non-trivial programs from these suites by filtering out input programs, where the
unmodified \TD\ has terminated in under five seconds.
We have extended the set of benchmarks with four open-source
Unix utilities \texttt{aget}, \texttt{ctrace}, \texttt{knot}, and \texttt{pfscan} to measure
performance on real-world applications.
For practical reasons, we have run the experiments
with a timeout of 20 minutes. This ruled out 24 programs where the unmodified \TD\ has timed out and one
further program where the \emph{independent} solver has timed out. In two cases, the joined solutions of
the solver threads from the \emph{independent} solver were, although sound, not fixpoints.
As this behavior leads to abnormal termination within the \wizwoz\ framework,
we have excluded these programs as well.

Our final benchmark consists of \numberTotalBenches\ programs (\numberSvBenches\ SV-Comp, \numberConratBenches\ Concrat, 4 Unix utilities) ranging from \minLoc\ to \maxLoc\ logical lines of code.
% We report the running-times for the respective solvers, not the complete analysis.
%
%
\ignore{
We also tried to find practical evidence for several of our algorithmic design decisions:
\begin{itemize}
    \item Is the CAS assumption of the \imme\ approach justified
    	  that collisions forcing repeated execution of operations are rare?
    \item \textbf{More to come!!!}
\end{itemize}
}
The experiments were conducted on Ubuntu 24.04.2 LTS with
two Intel(R) Xeon(R) Platinum 8260 CPUs, each with 24 cores, where a single core corresponds to the computing power of a core in an ordinary consumer machine.

% RQ1 - sequential improvements

\begin{description}
\item[\textbf{RQ1}.] We found that merging the various data structures of the original fixpoint algorithm, as
provided by \wizwoz\, in one hashmap \texttt{data} already resulted in a measurable speedup of
\refAvg $\times$ on average by reducing the time spent traversing and maintaining the data structures.
We observed that the runtime of fixpoint iteration is dependent on the execution order.
This is not surprising because if an unknown $x$ directly or indirectly depends on  another unknown $y$, calculating the value of $y$ before $x$ is faster, as $x$ must otherwise be recalculated.
Empirically, promoting the endpoints of called functions to the top-level workset at every call edge before querying their values resulted in a combined speedup in the
range of \baseFuncOneMin $\times$ to \baseFuncOneMax $\times$ with an average of \baseFuncOneAvg $\times$.
When unknowns were demanded at \verb|pthread_create| only, the speedup was even more pronounced, ranging from
\baseOneMin $\times$ to \baseOneMax $\times$ with an average of \baseOneAvg $\times$.
We conclude that the running time of \TD\ can be improved by iterating over certain unknowns on the top-level, where the granularity has a significant impact on the overall performance.

% Note for the editors: The lines in the following plots are not well distinguishable 
% in black and white. However, this is not central which concrete program behaves how.
% The imporant point is that the variation is visible, not which is which.
\begin{figure}[ht]
  \centering
  \begin{subfigure}[b]{0.48\textwidth}
    \includegraphics[width=\linewidth]{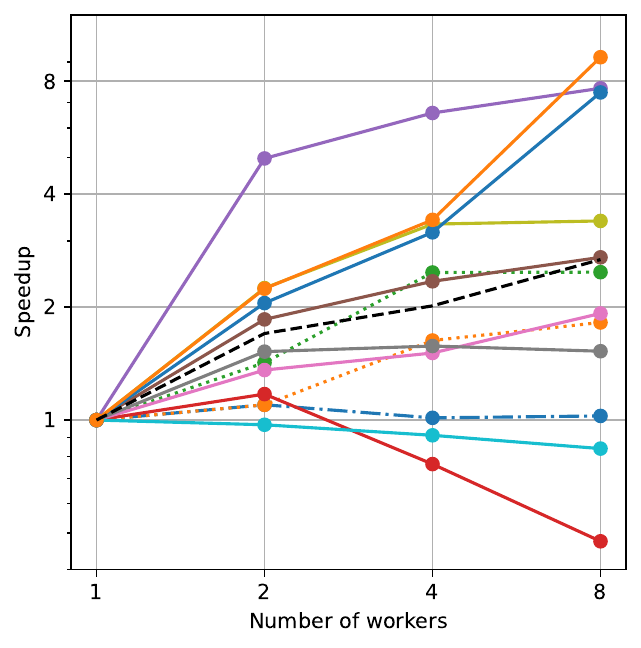}
    \caption{\imme\ approach speedups}
    \label{fig:1}
  \end{subfigure}
  \hfill
  \begin{subfigure}[b]{0.48\textwidth}
    \includegraphics[width=\linewidth]{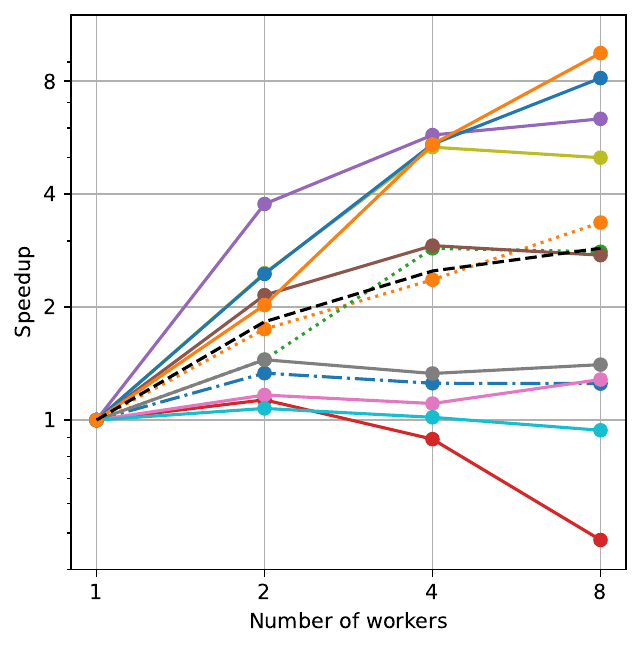}
    \caption{\indep\ approach speedups}
    \label{fig:2}
  \end{subfigure}
  \hfill
  \\
  \hfill
  \begin{subfigure}[b]{0.98\textwidth}
    \includegraphics[width=\linewidth]{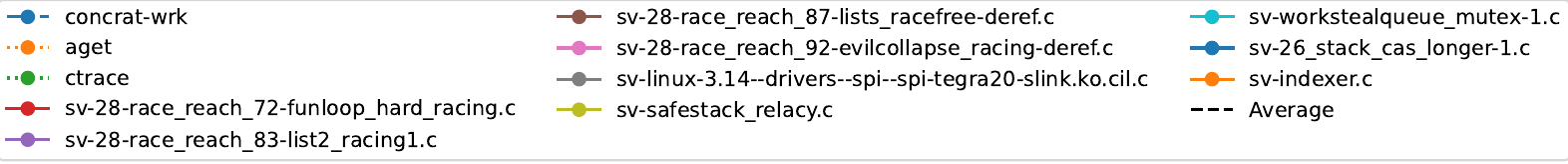}
    % \caption{Shared legend.}
  \end{subfigure}
  \caption{Relative speedups with demand at \texttt{pthread\_create}}
  \label{f:res:rq2}
\end{figure}

\begin{figure}[hbt]
\begin{center}
  \includegraphics[width=0.98\textwidth]{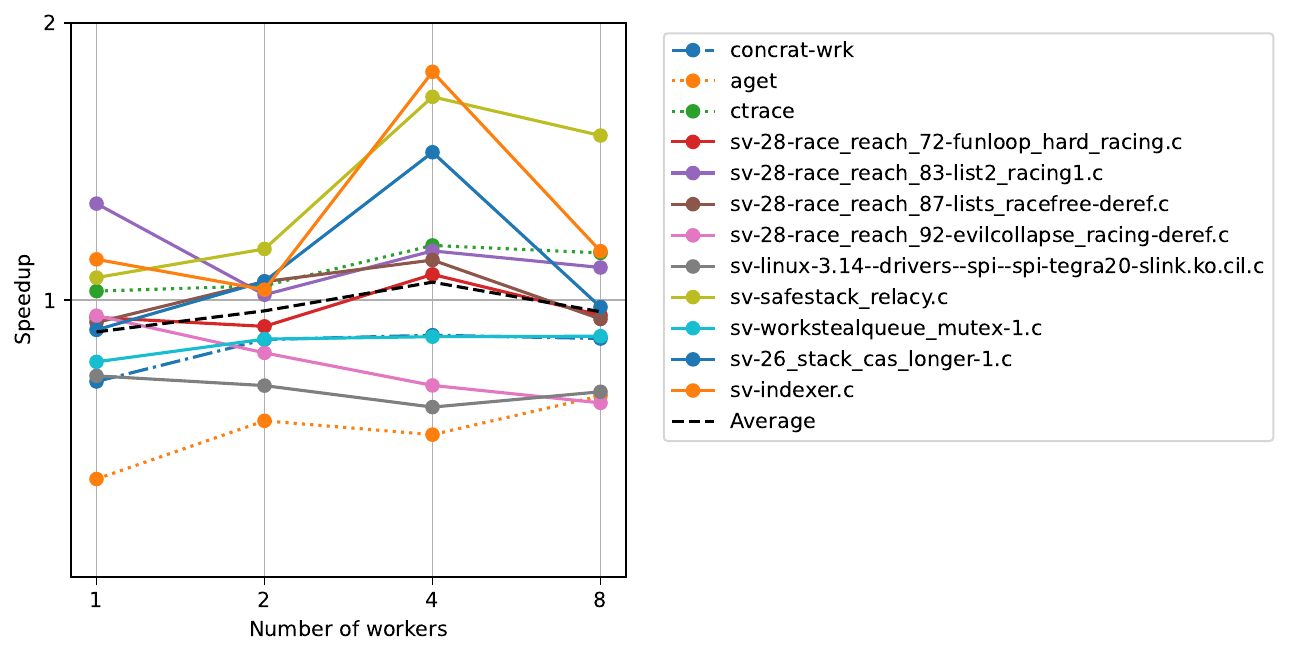}
  \caption{\label{f:res:rq2comp} Comparing \indep\ vs. \imme}
  \end{center}
\end{figure}

\item[\textbf{RQ2}.]
	We evaluated the two parallelizing approaches for 1, 2, 4, and 8 workers, where demands are placed at
\verb|pthread_create|.
For each benchmark program, we determine the sequence of speedups relative to the single-worker running time
of each approach.
Figure \ref{f:res:rq2} shows these speedups.
For better visualization of the results, we display the results only for a representative subset of programs
by selecting every \visualSamplingFactor\ program from the full benchmark suite, sorted by the single-worker running time.
Tables containing all running times
% for all benchmarks and configurations
can be found in the Appendix.

%
% each comparison with itself
%
The \imme\ approach on all benchmarks shows average (maximal) speed\-ups of \baseTwoSelfRelativeAvg $\times$ , \baseFourSelfRelativeAvg $\times$
and \baseEightSelfRelativeAvg $\times$ (\baseTwoSelfRelativeMax $\times$, \baseFourSelfRelativeMax $\times$, \baseEightSelfRelativeMax $\times$) with 2, 4 and 8 workers compared to when it is run with a single worker;
the \indep\ approach \distTwoSelfRelativeAvg $\times$, \distFourSelfRelativeAvg $\times$ and \distEightSelfRelativeAvg $\times$
(\distTwoSelfRelativeMax $\times$, \distFourSelfRelativeMax $\times$, \distEightSelfRelativeMax $\times$) respectively.
Thus, we find significant speedups consistently for both approaches, which increase with the number of workers. For the majority of our benchmark programs, speedups tend to diminish with more than four workers.
We observe super-linear speedups in some cases, which again is explained by the variations in execution order of the right-hand sides.
For a comparison of the two approaches,
Figure \ref{f:res:rq2comp} displays for every considered number of workers, the ratio of the running times of
the \imme\ over
the \indep\ solver in the same setting.
The independent and intermediate approaches have similar running-times, which almost never differ by a factor larger than 2. These differences are in either direction; neither approach dominates the other.
\begin{figure}[ht]
  \centering
  \begin{subfigure}[b]{0.48\textwidth}
    \includegraphics[width=\linewidth]{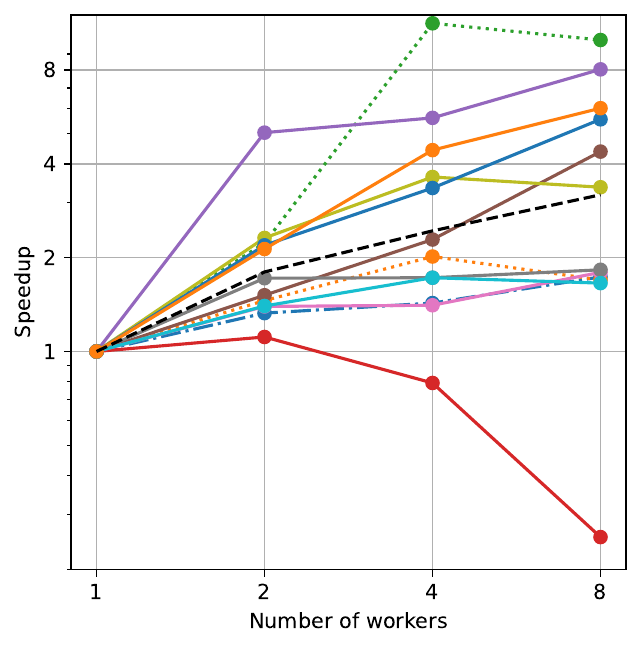}
    \caption{The \imme\ approach speedups with demand at functions}
  \end{subfigure}
  \hfill
  \begin{subfigure}[b]{0.48\textwidth}
    \includegraphics[width=\linewidth]{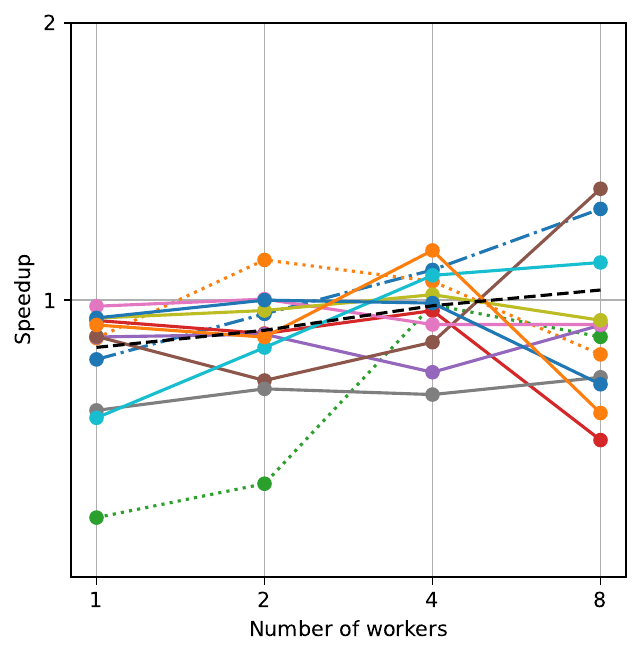}
    \caption{The \imme\ approach with demand at functions vs. \texttt{pthread\_create}}
  \end{subfigure}
  \hfill
  \\
  \hfill
  \begin{subfigure}[b]{0.98\textwidth}
    \includegraphics[width=\linewidth]{assets/legend.pdf}
    % \caption{Shared legend.}
  \end{subfigure}
  \caption{Putting demand at functions}
  \label{f:res:rq3}
\end{figure}

\item[\textbf{RQ3}.]
	To evaluate the impact of demand point selection, we have also run the \imme\ approach,
while placing demands not only when the analyzed program starts a new thread via \verb|pthread_create|, but also for the unknowns corresponding to the endpoints of every called function.
Figure \ref{f:res:rq3} displays these results in the same way as above.
In \textbf{RQ1}, we noticed that using demand nodes sparingly, only for \verb!pthread! creation,
leads to better \emph{single-worker} performance. In contrast,
we find that the increased number of roots results in better scaling with the number of workers.
% and in \todo{how many?} many cases, better overall performance.
This indicates that the potential for parallelism depends on the selection of demand points.
In the input programs where the introduced potential for parallelism by the mentioned strategies was not large enough, speedups diminished
with higher parallelization overhead incurred by larger numbers of solver threads.
\item[\textbf{RQ4}.] As expected, the results computed by the various solver configurations are not always
identical.
Still, we generally found agreement in the results for the vast majority of unknowns (see Table \ref{t:precision}).
On average, slightly more unknowns receive more rather than less precise values compared to the
solution by the unmodified \TD. Barely any unknowns have incomparable values.

\setlength{\tabcolsep}{8pt} % increase spacing a bit
\renewcommand{\arraystretch}{1.2} % add vertical padding

\begin{table}[ht]
\centering
\begin{tabular}{@{\hskip 3pt}l@{\hskip 8pt}l@{\hskip 8pt}l@{\hskip 8pt}l@{\hskip 8pt}l@{\hskip 3pt}}
\toprule
\textbf{Approach} & \textbf{Demand} & \textbf{Equal} & \textbf{More precise} & \textbf{Less precise} \\
\midrule
\imme  & pthread            & \baseSolverEqualPrecisionRate     & \baseSolverBetterPrecisionRate     & \baseSolverWorsePrecisionRate \\
\imme  & pthread, functions & \baseFuncSolverEqualPrecisionRate & \baseFuncSolverBetterPrecisionRate & \baseFuncSolverWorsePrecisionRate \\
\indep & pthread            & \distSolverEqualPrecisionRate     & \distSolverBetterPrecisionRate     & \distSolverWorsePrecisionRate \\
\bottomrule
\end{tabular}
\caption{Precision across approaches, compared with unmodified \TD.}
\label{t:precision}
\end{table}
\end{description}

\paragraph{Threats to Validity.}
A fundamental issue is how far our collection of benchmarks is representative of relevant applications.
By including the non-trivial programs from already proposed suites, namely, the non-trivial concurrency benchmarks
from SV-Comp and the Concrat suite, we included examples that have already been judged as reasonable by the
community.
By adding further multi-threaded open source Unix utilities, we tried to complement
these existing suites with further commonly used programs.
Indeed, our findings for programs are quite independent of their sources.
When reporting our results for runs of the parallel solvers, we must be aware that, due to the operating system's scheduling policy or the \textsc{OCaml} runtime's memory management, non-deterministic behavior is to be expected.
We have reported the number of single runs here to expose irregularities in performance and precision.
We observe, however, that the averaged results across three runs, 
both for running times and the ratio of unknowns with equal precision, vary by less than 2\%.

Another issue is our choice of the analysis configuration. To obtain comparable results,
we did not report on results with different abstract domains or other approaches to distinguish multiple
calling contexts. Our experiments, though, indicate that the outcome for these is essentially
the same.
Finally, it is difficult to directly compare experimental results with state-of-the-art parallelization
approaches, as the underlying analyzers often have fundamental differences in their requirements and
limitations. Further, the approaches we consider to have the most similarities are implemented for analyzers
of different programming languages, making it unfeasible to evaluate them on the same benchmark suite.

\section{Related Work}\label{s:work}
Parallelization to speed up program analysis tasks is widely applied to 
model checking and dynamic or static analyses.
We focus on static analysis and specifically on techniques that parallelize the analyzer itself,
as opposed to, e.g., \emph{parallel portfolio} approaches where multiple (configurations of)
analyzers are run in parallel~\cite{beyerportfolio,nacpa,huntingparty},
% could also talk about other parallelization approaches (splitting etc), see Marie-Christine's paper
% (https://arxiv.org/pdf/2509.13699), but maybe in the interest of space we skip that?
or approaches that compile 
analysis specifications into \textsc{Datalog} programs and then use concurrent evaluation engines~(e.g., \cite{flan,souflee,ascent,jordanParallelDatalog}).

Analyses that compute bottom-up summaries of procedures are
easy to parallelize, as all leaves can be analyzed in parallel. \citet{pipliningbottomup}
report that many analyzers~\cite{XieAiken2005,BabicHu2008,ShiZXZFF18,Infer,scalableandincremental} 
make use of this idea.
Based on this observation, \citet{pipliningbottomup} propose to increase parallelism
by decomposing the analysis of a single procedure into multiple tasks.
% : as the state up-to a function
% call for which a summary is not yet available can be computed in parallel to this summary.

\citet{parallelAstree} parallelizes the \textsc{Astrée} analyzer by computing abstract states for both branches for
long-running \emph{if-then-else} statements in parallel. In particular, he focuses on \emph{dispatch} statements that
occur in event-driven programs, where, based on the event to be processed, different potentially long-running
code paths are taken.
\citet{pikos} parallelize the \textsc{Ikos} abstract interpreter by further
weakening Bourdoncle's weak topological order~\cite{bourdoncle1993efficient} to a
weak \emph{partial} order~(wpo), allowing parallelization of parts of the program
that are unordered w.r.t.\ each other. They also establish that
their algorithm is deterministic: any scheduling choice consistent with the
wpo will compute the same result as Bourdoncle's original algorithm.
These approaches expect the complete equation system be provided explicitly beforehand ---
which is not the case for context-sensitive and especially demand-driven analyses, where
an implicitly provided equation system is only partly explored.
Approaches requiring preprocessing of the equation system, therefore, are not applicable.

The two strands of work most closely related to ours are by~\citet{bolt} and \citet{modularworklist}
(later presented in an extended version~\cite{worklistlong}), respectively.
Both approaches require a set of predefined modules (functions, and additionally threads in the latter work).
In our setting, however, such predefined modules are not readily available as we target the parallelization
of \emph{generic} solvers.
%
% for arbitrary side-effecting constraint systems.
%
\citet{modularworklist} treat the worklist as a task pool of components to be analyzed. The analysis
of each such component yields new components to be analyzed (either because they are newly discovered
or because they need to be re-analyzed as they depend on the results of the current component).
In this way, communication of updates to dependencies happens only when the analysis of a component
has finished, whereas in our approaches, updates to globals are communicated between tasks continuously.
% \todo{SS: Are we claiming to be better then or just different?}
%
\citet{bolt}, on the other hand, take a more structured approach to concurrency and alternate between
\emph{map} and \emph{reduce} phases. In the \emph{map} phase, queries in the worklist are processed until 
they yield results or are blocked on dependencies. In the \emph{reduce} phase, the state of all queries is 
updated, and queries that are unblocked are added back to the worklist.
In our approaches, tasks do not become blocked when querying a top-level unknown that is still being
computed by another task. Instead, they receive the current best knowledge about that unknown and continue
their work optimistically, but they will be rescheduled if the unknown is later updated.

% \todo[inline]{The following are not-too-closely related and should be condensed or removed if we need space.}
\citet{actorifds} parallelize IFDS-style analyses \cite{reps} by defining one actor per node and realizing 
propagation of dataflow facts as message passing between actors, delegating 
the propagation of messages to the runtime system.
%
% This relies on the structure of the equation system being known ahead-of-time. 
% It is not clear how this approach could be extended to demand-driven analyses.
% \todo{Could also cite Bodden and his Heros solver?}
\citet{semiimplicit} also achieve parallelization of IFDS-style analyses,
using a \emph{semi-implicit} programming model that propagates values between \emph{cells} where
developers explicitly give dependencies between cells and mark some cells as needing
to be updated non-concurrently. They require monotone right-hand sides and
bounded lattices satisfying the ascending chain condition, and do not support widening and narrowing.

\citet{bigdataflow} propose to execute dataflow analyses given an interprocedural control-flow graph
on large compute clusters based on Apache Giraph. To this end, they devise a \emph{distributed} fixpoint algorithm 
tuned to reduce the amount of communication between compute nodes by only propagating new dataflow facts, 
akin to \emph{semi-naïve evaluation}.
This approach is once more \emph{not} applicable to demand-driven analyses and does not support 
widening and narrowing.
% Nevertheless, it would be interesting to see how our techniques could be combined
% with their work to go from a parallel to a distributed setting.

\citet{cdfa} propose to parallelize the analysis at a \emph{fine-grained} level by splitting up CFGs and 
considering some paths in the CFG in parallel. Once more, this approach is not applicable to demand-driven analyses
with dynamic dependencies between unknowns.

\citet{ParallelJS} take a different approach and present the analysis of \textsc{Java\-Script} programs as an 
embarrassingly parallel problem where all paths are analyzed in parallel, with a separate process controlling 
which abstract program states are to be merged. The motivation is that in highly dynamic languages such as 
\textsc{JavaScript}, CFGs are not readily available. This kind of forward exploration does not easily extend to 
the demand-driven analyses, which we study here.
% They have different strategies where they parallelize the entire worklist, and one they call per-context where
% a thread is associated with a context and all work in this context is done by this thread.

\section{Conclusion and Future Work}\label{s:final}

We have presented two approaches for parallelizing the generic fixpoint engine \TTD\
and have demonstrated their effectiveness for several gears, i.e., granularities,
in the context of a thread-modular analysis
as provided by the static analyzer \wizwoz.
Already, the modifications to the original \TD\ from \wizwoz\ in preparation for the
parallelization yielded speedups up to a factor of 10, with an average speedup
of \baseFuncOneAvg$\times$.
When turning to parallelization, we observed \emph{further} speedups of up to a factor
of \baseFuncFourSelfRelativeMax$\times$ (4 workers, \imme\ approach, demand at functions),
respectively \distEightSelfRelativeMax$\times$ (8 workers, \indep\ approach, demands at thread creations),
with an average speedup of \baseFuncEightSelfRelativeAvg$\times$ (\imme, 8 workers) and \distEightSelfRelativeAvg$\times$ (\indep, 8 workers).
The two parallelization approaches complement each other with no clear winner.
It seems, though, that for smaller granularity, e.g., at the procedure level, the
\imme\ approach is preferable.

Here, we have presented results only from experiments with two granularities.
It remains for future work to explore further possibilities of $\demand$ placement,
depending, e.g., on the characteristics of the program under analysis.

\paragraph*{Acknowledgments.}
This work was supported in part by
the European Union and the Estonian Research Council via projects PRG2764 and TEM-TA119.
This work has benefited from the participation of Helmut Seidl and Michael Schwarz
in the Dagstuhl Seminar 25421 \textit{Sound Static Program Analysis in Modern Software Engineering}.

\paragraph*{Data-Availability Statement}
The artifact, including source code for our implementations, benchmarks, benchmark results and the evaluation scripts,
is archived and available at DOI \href{https://doi.org/10.5281/zenodo.18186778}{10.5281/zenodo.18186778}\cite{artifact}.

\bibliographystyle{splncs04nat}
\bibliography{bibliography}
%

% LNCS Homepage, \url{http://www.springer.com/lncs}, last accessed 2023/10/25
% \end{thebibliography}
\ifdefined\extended
\appendix
\newpage
\section{Appendix: All Running-times}\label{a:runtimes}

The following tables are sorted by the running time of \TD\ without any modification.

% \begin{table}[ht]
\begin{longtable}{p{8cm}@{\hskip 8pt}r@{\qquad}r}
\caption{Running-times of all benchmarks with \TD\ and \TTD\ (without any demands). The difference thus is due
purely to the optimized data structures, and not the order of evaluation.}\label{tab:allruntimes}\\
\toprule
file & \TD & \TTD\\
\midrule
sv-28-race\_reach\_87-lists\_racefree-deref.c & 5.02 & 4.42 \\
sv-28-race\_reach\_87-lists\_racefree.c & 5.09 & 4.43 \\
sv-28-race\_reach\_86-lists\_racing.c & 5.11 & 4.53 \\
sv-28-race\_reach\_94-alloc\_region\_racing.c & 5.19 & 3.44 \\
sv-safestack\_relacy.c & 5.59 & 4.93 \\
sv-26\_stack\_cas\_longest-1-race.c & 5.77 & 4.54 \\
sv-26\_stack\_cas\_longer-1-race.c & 5.80 & 4.57 \\
sv-26\_stack\_cas\_longest-1.c & 6.30 & 5.07 \\
sv-26\_stack\_cas\_longer-1.c & 6.31 & 5.04 \\
sv-09-regions\_17-arrayloop\_nr.c & 6.46 & 5.73 \\
sv-linux-3.14--drivers--usb--misc--adutux.ko.cil.c & 6.69 & 5.01 \\
concrat-snoopy-mini & 6.86 & 6.76 \\
sv-28-race\_reach\_83-list2\_racing1.c & 8.19 & 6.75 \\
sv-28-race\_reach\_85-list2\_racefree.c & 8.35 & 7.00 \\
sv-workstealqueue\_mutex-2.c & 9.01 & 8.15 \\
concrat-snoopy & 9.14 & 8.71 \\
sv-workstealqueue\_mutex-1.c & 9.31 & 8.35 \\
concrat-EasyLogger-mini & 14.45 & 11.84 \\
sv-28-race\_reach\_84-list2\_racing2.c & 16.12 & 13.80 \\
concrat-uthash & 19.99 & 16.92 \\
sv-linux-3.14--drivers--spi--spi-tegra20-slink.ko.cil.c & 29.85 & 22.67 \\
concrat-libaco & 36.54 & 32.21 \\
concrat-vanitygen & 42.70 & 33.85 \\
sv-linux-3.14--drivers--net--irda--w83977af\_ir.ko.cil.c & 45.42 & 27.99 \\
sv-28-race\_reach\_92-evilcollapse\_racing-deref.c & 48.62 & 44.35 \\
sv-28-race\_reach\_91-arrayloop2\_racefree.c & 49.00 & 45.47 \\
sv-28-race\_reach\_93-evilcollapse\_racefree.c & 50.05 & 45.23 \\
sv-elimination\_backoff\_stack-race.c & 68.74 & 63.14 \\
sv-indexer.c & 77.24 & 51.19 \\
sv-28-race\_reach\_92-evilcollapse\_racing.c & 84.81 & 81.33 \\
sv-elimination\_backoff\_stack.c & 89.40 & 83.18 \\
pfscan & 103.25 & 97.45 \\
ctrace & 124.02 & 105.27 \\
concrat-pingfs & 163.10 & 102.96 \\
concrat-nnn-mini & 179.71 & 149.25 \\
concrat-stud & 193.68 & 175.32 \\
aget & 193.97 & 182.34 \\
sv-28-race\_reach\_90-arrayloop2\_racing.c & 202.13 & 192.71 \\
knot & 226.51 & 198.67 \\
concrat-ProcDump-for-Linux-mini & 233.36 & 232.03 \\
concrat-wrk & 244.01 & 179.00 \\
sv-28-race\_reach\_70-funloop\_racefree.c & 414.92 & 365.53 \\
concrat-EasyLogger & 445.13 & 411.45 \\
sv-28-race\_reach\_73-funloop\_hard\_racefree.c & 454.80 & 419.66 \\
sv-28-race\_reach\_72-funloop\_hard\_racing.c & 469.78 & 417.31 \\
concrat-cava & 479.46 & 458.32 \\
sv-28-race\_reach\_71-funloop\_racing.c & 543.79 & 480.52 \\
\bottomrule
\end{longtable}

% \begin{tabular}{lrrrr}
  \begin{longtable}{l@{\hskip 0pt}r@{\hskip 3pt}r@{\hskip 3pt}r@{\hskip 3pt}r}
    \caption{Running-times of all benchmarks with the \emph{immediate} solver with demands at \texttt{pthread\_create}, by number of workers}\label{tab:allruntimesimmediatiate}\\
\toprule
file &  1 & 2 & 4 & 8 \\
\midrule
sv-28-race\_reach\_87-lists\_racefree-deref.c & 4.45 & 2.40 & 1.90 & 1.64 \\
sv-28-race\_reach\_87-lists\_racefree.c & 4.44 & 2.29 & 1.97 & 1.20 \\
sv-28-race\_reach\_86-lists\_racing.c & 4.48 & 2.58 & 2.20 & 1.38 \\
sv-28-race\_reach\_94-alloc\_region\_racing.c & 1.42 & 1.20 & 1.11 & 1.02 \\
sv-safestack\_relacy.c & 1.73 & 0.77 & 0.52 & 0.51 \\
sv-26\_stack\_cas\_longest-1-race.c & 2.63 & 1.28 & 0.77 & 0.33 \\
sv-26\_stack\_cas\_longer-1-race.c & 2.61 & 1.29 & 0.80 & 0.36 \\
sv-26\_stack\_cas\_longest-1.c & 2.86 & 1.50 & 0.73 & 0.42 \\
sv-26\_stack\_cas\_longer-1.c & 2.91 & 1.42 & 0.92 & 0.39 \\
sv-09-regions\_17-arrayloop\_nr.c & 1.87 & 2.09 & 2.47 & 2.33 \\
sv-linux-3.14--drivers--usb--misc--adutux.ko.cil.c & 4.53 & 2.96 & 2.47 & 2.62 \\
concrat-snoopy-mini & 2.86 & 1.98 & 1.27 & 1.21 \\
sv-28-race\_reach\_83-list2\_racing1.c & 13.09 & 2.63 & 1.99 & 1.71 \\
sv-28-race\_reach\_85-list2\_racefree.c & 10.38 & 2.69 & 2.35 & 1.56 \\
sv-workstealqueue\_mutex-2.c & 5.88 & 6.39 & 6.20 & 7.57 \\
concrat-snoopy & 4.82 & 3.08 & 2.50 & 1.96 \\
sv-workstealqueue\_mutex-1.c & 6.11 & 6.29 & 6.71 & 7.28 \\
concrat-EasyLogger-mini & 6.14 & 2.08 & 2.16 & 2.27 \\
sv-28-race\_reach\_84-list2\_racing2.c & 15.98 & 3.94 & 3.36 & 2.70 \\
concrat-uthash & 18.96 & 11.98 & 14.90 & 16.13 \\
sv-linux-3.14--drivers--spi--spi-tegra20-slink.ko.cil.c & 11.70 & 7.69 & 7.43 & 7.67 \\
concrat-libaco & 15.79 & 6.42 & 6.59 & 7.19 \\
concrat-vanitygen & 36.36 & 38.39 & 39.90 & 39.32 \\
sv-linux-3.14--drivers--net--irda--w83977af\_ir.ko.cil.c & 19.28 & 13.22 & 13.01 & 13.77 \\
sv-28-race\_reach\_92-evilcollapse\_racing-deref.c & 26.54 & 19.53 & 17.59 & 13.80 \\
sv-28-race\_reach\_91-arrayloop2\_racefree.c & 27.00 & 19.16 & 17.76 & 17.96 \\
sv-28-race\_reach\_93-evilcollapse\_racefree.c & 27.45 & 19.39 & 16.39 & 15.64 \\
sv-elimination\_backoff\_stack-race.c & 35.36 & 21.30 & 25.98 & 29.96 \\
sv-indexer.c & 7.41 & 3.31 & 2.17 & 0.80 \\
sv-28-race\_reach\_92-evilcollapse\_racing.c & 47.03 & 33.84 & 32.69 & 35.50 \\
sv-elimination\_backoff\_stack.c & 53.05 & 29.26 & 36.44 & 43.46 \\
pfscan & 25.12 & 14.57 & 11.36 & 8.82 \\
ctrace & 36.62 & 25.76 & 14.81 & 14.78 \\
concrat-pingfs & 62.66 & 65.88 & 62.76 & 62.45 \\
concrat-nnn-mini & 53.94 & 72.06 & 100.21 & 102.74 \\
concrat-stud & 139.74 & 130.84 & 120.08 & 123.34 \\
aget & 96.06 & 87.52 & 58.97 & 52.80 \\
sv-28-race\_reach\_90-arrayloop2\_racing.c & 117.97 & 71.87 & 61.94 & 54.56 \\
knot & 211.21 & 136.87 & 79.01 & 43.39 \\
concrat-ProcDump-for-Linux-mini & 248.43 & 131.16 & 106.85 & 113.44 \\
concrat-wrk & 121.71 & 110.83 & 120.09 & 118.68 \\
sv-28-race\_reach\_70-funloop\_racefree.c & 137.00 & 137.41 & 205.86 & 365.11 \\
concrat-EasyLogger & 198.02 & 136.22 & 184.98 & 164.50 \\
sv-28-race\_reach\_73-funloop\_hard\_racefree.c & 137.94 & 129.24 & 202.67 & 387.21 \\
sv-28-race\_reach\_72-funloop\_hard\_racing.c & 149.75 & 127.71 & 196.23 & 315.09 \\
concrat-cava & 800.08 & 494.38 & 467.60 & 491.69 \\
sv-28-race\_reach\_71-funloop\_racing.c & 144.95 & 143.62 & 221.97 & 302.83 \\
\bottomrule
\end{longtable}

  \begin{longtable}{l@{\hskip 0pt}r@{\hskip 1pt}r@{\hskip 1pt}r@{\hskip 1pt}r}
    \caption{Running-times of all benchmarks with the \emph{immediate} solver with demands at functions, by number of workers}\label{tab:allruntimesimmediatefunc}\\
\toprule
file &  1 & 2 & 4 & 8 \\
\midrule
sv-28-race\_reach\_87-lists\_racefree-deref.c & 5.12 & 3.38 & 2.24 & 1.17 \\
sv-28-race\_reach\_87-lists\_racefree.c & 5.35 & 3.01 & 2.14 & 1.16 \\
sv-28-race\_reach\_86-lists\_racing.c & 4.81 & 3.28 & 2.32 & 1.19 \\
sv-28-race\_reach\_94-alloc\_region\_racing.c & 1.49 & 1.34 & 1.09 & 0.94 \\
sv-safestack\_relacy.c & 1.85 & 0.80 & 0.51 & 0.55 \\
sv-26\_stack\_cas\_longest-1-race.c & 2.88 & 1.32 & 0.73 & 0.30 \\
sv-26\_stack\_cas\_longer-1-race.c & 2.90 & 1.24 & 0.82 & 0.32 \\
sv-26\_stack\_cas\_longest-1.c & 3.10 & 1.66 & 0.72 & 0.39 \\
sv-26\_stack\_cas\_longer-1.c & 3.11 & 1.42 & 0.93 & 0.56 \\
sv-09-regions\_17-arrayloop\_nr.c & 3.19 & 2.27 & 1.77 & 1.40 \\
sv-linux-3.14--drivers--usb--misc--adutux.ko.cil.c & 5.84 & 3.38 & 2.59 & 3.05 \\
concrat-snoopy-mini & 2.74 & 1.72 & 1.33 & 1.00 \\
sv-28-race\_reach\_83-list2\_racing1.c & 15.10 & 3.00 & 2.69 & 1.88 \\
sv-28-race\_reach\_85-list2\_racefree.c & 11.25 & 3.18 & 2.14 & 2.06 \\
sv-workstealqueue\_mutex-2.c & 9.75 & 7.51 & 6.88 & 6.12 \\
concrat-snoopy & 4.70 & 3.01 & 2.07 & 1.74 \\
sv-workstealqueue\_mutex-1.c & 10.62 & 7.59 & 6.16 & 6.41 \\
concrat-EasyLogger-mini & 6.43 & 2.18 & 2.19 & 2.30 \\
sv-28-race\_reach\_84-list2\_racing2.c & 18.74 & 5.28 & 4.45 & 2.75 \\
concrat-uthash & 19.13 & 12.00 & 14.03 & 13.58 \\
sv-linux-3.14--drivers--spi--spi-tegra20-slink.ko.cil.c & 19.43 & 11.31 & 11.27 & 10.63 \\
concrat-libaco & 17.32 & 6.91 & 6.49 & 6.54 \\
concrat-vanitygen & 39.33 & 26.35 & 30.60 & 26.22 \\
sv-linux-3.14--drivers--net--irda--w83977af\_ir.ko.cil.c & 39.58 & 26.67 & 24.26 & 17.23 \\
sv-28-race\_reach\_92-evilcollapse\_racing-deref.c & 27.14 & 19.46 & 19.29 & 15.13 \\
sv-28-race\_reach\_91-arrayloop2\_racefree.c & 27.87 & 18.58 & 18.07 & 16.08 \\
sv-28-race\_reach\_93-evilcollapse\_racefree.c & 28.37 & 19.52 & 18.49 & 13.31 \\
sv-elimination\_backoff\_stack-race.c & 55.09 & 26.49 & 26.53 & 30.81 \\
sv-indexer.c & 8.14 & 3.82 & 1.84 & 1.35 \\
sv-28-race\_reach\_92-evilcollapse\_racing.c & 48.00 & 34.37 & 31.69 & 30.86 \\
sv-elimination\_backoff\_stack.c & 83.26 & 45.62 & 37.47 & 52.14 \\
pfscan & 25.43 & 15.22 & 11.71 & 8.66 \\
ctrace & 169.98 & 76.28 & 15.06 & 17.04 \\
concrat-pingfs & 81.56 & 67.92 & 63.11 & 73.40 \\
concrat-nnn-mini & 144.00 & 274.39 & 102.30 & 32.18 \\
concrat-stud & 155.47 & 136.76 & 132.53 & 121.25 \\
aget & 111.33 & 76.44 & 55.22 & 65.66 \\
sv-28-race\_reach\_90-arrayloop2\_racing.c & 120.50 & 69.41 & 51.86 & 42.86 \\
knot & 226.80 & 149.80 & 90.93 & 59.48 \\
concrat-ProcDump-for-Linux-mini & 323.52 & 156.11 & 151.45 & 152.56 \\
concrat-wrk & 154.83 & 116.62 & 108.31 & 89.30 \\
sv-28-race\_reach\_70-funloop\_racefree.c & 160.45 & 147.62 & 226.91 & 758.70 \\
concrat-EasyLogger & 238.74 & 148.14 & 154.16 & 152.31 \\
sv-28-race\_reach\_73-funloop\_hard\_racefree.c & 162.47 & 149.13 & 226.44 & 583.89 \\
sv-28-race\_reach\_72-funloop\_hard\_racing.c & 161.62 & 145.22 & 203.96 & 636.51 \\
concrat-cava & 1178.85 & 522.54 & 492.47 & 478.89 \\
sv-28-race\_reach\_71-funloop\_racing.c & 164.79 & 145.31 & 232.06 & 499.13 \\
\bottomrule
\end{longtable}

 \begin{longtable}{l@{\hskip 0pt}r@{\hskip 3pt}r@{\hskip 3pt}r@{\hskip 3pt}r}
    \caption{Running-times of all benchmarks with the \emph{independent} solver with demands at \texttt{pthread\_create}, by number of workers}\label{tab:allruntimesindependent}\\
\toprule
file &  1 & 2 & 4 & 8 \\
\midrule
sv-28-race\_reach\_87-lists\_racefree-deref.c & 4.84 & 2.25 & 1.66 & 1.76 \\
sv-28-race\_reach\_87-lists\_racefree.c & 4.81 & 2.40 & 1.57 & 1.91 \\
sv-28-race\_reach\_86-lists\_racing.c & 4.85 & 3.12 & 1.86 & 1.87 \\
sv-28-race\_reach\_94-alloc\_region\_racing.c & 1.14 & 0.61 & 0.70 & 0.57 \\
sv-safestack\_relacy.c & 1.60 & 0.65 & 0.30 & 0.32 \\
sv-26\_stack\_cas\_longest-1-race.c & 2.93 & 0.69 & 0.38 & 0.32 \\
sv-26\_stack\_cas\_longer-1-race.c & 2.94 & 0.80 & 0.40 & 0.33 \\
sv-26\_stack\_cas\_longest-1.c & 3.23 & 1.42 & 0.61 & 0.41 \\
sv-26\_stack\_cas\_longer-1.c & 3.26 & 1.33 & 0.60 & 0.40 \\
sv-09-regions\_17-arrayloop\_nr.c & 1.91 & 2.12 & 2.56 & 2.46 \\
sv-linux-3.14--drivers--usb--misc--adutux.ko.cil.c & 4.23 & 3.52 & 2.46 & 2.85 \\
concrat-snoopy-mini & 5.25 & 3.96 & 2.10 & 1.61 \\
sv-28-race\_reach\_83-list2\_racing1.c & 9.71 & 2.58 & 1.69 & 1.53 \\
sv-28-race\_reach\_85-list2\_racefree.c & 10.74 & 2.72 & 2.22 & 1.57 \\
sv-workstealqueue\_mutex-2.c & 7.83 & 7.33 & 7.81 & 7.97 \\
concrat-snoopy & 7.10 & 4.26 & 3.17 & 2.17 \\
sv-workstealqueue\_mutex-1.c & 7.86 & 7.32 & 7.73 & 8.37 \\
concrat-EasyLogger-mini & 9.56 & 2.96 & 3.07 & 3.17 \\
sv-28-race\_reach\_84-list2\_racing2.c & 16.02 & 5.08 & 3.08 & 2.63 \\
concrat-uthash & 16.79 & 10.32 & 10.87 & 11.47 \\
sv-linux-3.14--drivers--spi--spi-tegra20-slink.ko.cil.c & 16.10 & 11.12 & 12.10 & 11.46 \\
concrat-libaco & 12.96 & 5.01 & 5.23 & 5.29 \\
concrat-vanitygen & 34.97 & 36.41 & 37.77 & 38.83 \\
sv-linux-3.14--drivers--net--irda--w83977af\_ir.ko.cil.c & 36.10 & 31.06 & 34.31 & 29.77 \\
sv-28-race\_reach\_92-evilcollapse\_racing-deref.c & 28.12 & 24.12 & 25.41 & 21.93 \\
sv-28-race\_reach\_91-arrayloop2\_racefree.c & 29.43 & 23.50 & 22.64 & 22.11 \\
sv-28-race\_reach\_93-evilcollapse\_racefree.c & 28.81 & 21.07 & 22.63 & 22.69 \\
sv-elimination\_backoff\_stack-race.c & 33.00 & 17.29 & 15.30 & 16.77 \\
sv-indexer.c & 6.46 & 3.19 & 1.19 & 0.68 \\
sv-28-race\_reach\_92-evilcollapse\_racing.c & 47.73 & 36.83 & 41.05 & 34.02 \\
sv-elimination\_backoff\_stack.c & 50.17 & 21.68 & 22.38 & 37.59 \\
pfscan & 23.41 & 16.94 & 10.64 & 8.07 \\
ctrace & 35.47 & 24.49 & 12.37 & 12.63 \\
concrat-pingfs & 101.64 & 105.06 & 108.10 & 112.63 \\
concrat-nnn-mini & 113.09 & 142.58 & 155.05 & 167.36 \\
concrat-stud & 143.26 & 118.47 & 126.43 & 119.04 \\
aget & 270.91 & 155.04 & 114.58 & 80.65 \\
sv-28-race\_reach\_90-arrayloop2\_racing.c & 130.57 & 87.97 & 63.62 & 88.51 \\
knot & 351.63 & 201.23 & 128.36 & 71.19 \\
concrat-ProcDump-for-Linux-mini & 411.38 & 185.67 & 165.17 & 166.32 \\
concrat-wrk & 172.30 & 129.22 & 137.50 & 137.78 \\
sv-28-race\_reach\_70-funloop\_racefree.c & 152.24 & 121.79 & 185.16 & 321.56 \\
concrat-EasyLogger & 350.83 & 210.12 & 221.65 & 244.90 \\
sv-28-race\_reach\_73-funloop\_hard\_racefree.c & 157.23 & 142.04 & 172.59 & 328.60 \\
sv-28-race\_reach\_72-funloop\_hard\_racing.c & 159.72 & 141.08 & 179.58 & 333.34 \\
concrat-cava & 850.30 & 389.75 & 413.46 & 404.21 \\
sv-28-race\_reach\_71-funloop\_racing.c & 159.44 & 145.27 & 203.60 & 381.46 \\
\bottomrule
\end{longtable}

\fi

\end{document}